\documentclass[twocolumn,aps,amsmath,amssymb,floatfix,superscriptaddress,prb,longbibliography]{revtex4-1}
\usepackage{graphicx}
\usepackage{dcolumn}
\usepackage{bm}
\usepackage{xspace}
\usepackage{hyperref}
\usepackage{color}
\usepackage[flushleft]{threeparttable}

\def\kyvo{K$_3$Yb(VO$_4$)$_2$\xspace}
\def\ryvo{Rb$_3$Yb(VO$_4$)$_2$\xspace}
\def\cyvo{Cs$_3$Yb(VO$_4$)$_2$\xspace}
\begin{document}

\title{Magnetically disordered ground state in the triangular-lattice antiferromagnets \ryvo and \cyvo}

\author{Zhen~Ma}
\email{zma@hbnu.edu.cn}
\author{Yingqi~Chen}
\author{Zhongtuo~Fu}
\author{Shuaiwei~Li}
\author{Xin-An~Tong}
\affiliation{Hubei Key Laboratory of Photoelectric Materials and Devices, School of Materials Science and Engineering, Hubei Normal University, Huangshi 435002, China}
\author{Hong~Du}
\affiliation{Tsung-Dao Lee Institute $\&$ School of Physics and Astronomy, Shanghai Jiao Tong University, Shanghai 200240, China}
\author{Jan Peter Embs}
\affiliation{PSI Center for Neutron and Muon Sciences, 5232 Villigen PSI, Switzerland}
\author{Shuhan~Zheng}
\author{Yongjun~Zhang}
\author{Meifeng~Liu}
\affiliation{Hubei Key Laboratory of Photoelectric Materials and Devices, School of Materials Science and Engineering, Hubei Normal University, Huangshi 435002, China}
\author{Ruidan~Zhong}
\affiliation{Tsung-Dao Lee Institute $\&$ School of Physics and Astronomy, Shanghai Jiao Tong University, Shanghai 200240, China}
\author{Jun-Ming~Liu}
\author{Jinsheng~Wen}
\email{jwen@nju.edu.cn}
\affiliation{National Laboratory of Solid State Microstructures and Department of Physics, Nanjing University, Nanjing 210093, China}
\affiliation{Collaborative Innovation Center of Advanced Microstructures and Jiangsu Physical Science Research Center, Nanjing University, Nanjing 210093, China}


\begin{abstract}
Quantum spin liquids~(QSLs) represent a unique quantum disordered state of matter that hosts long-range quantum entanglement and fractional excitations. However, structural disorder resulting from site mixing between different types of ions usually arises in real QSL candidates, which is considered as an obstacle to gain the insight into the intrinsic physics. Here, we have synthesized two new rare-earth compounds \ryvo and \cyvo. X-ray diffractions reveal a perfect triangular-lattice structure with no detectable disorder. Magnetic susceptibility measurements do not capture any phase transition or spin freezing down to 1.8~K. A fit to low-temperature data indicates dominant antiferromagnetic interactions with the Curie-Weiss temperature of -1.40~K and -0.43~K for \ryvo and \cyvo, respectively. Specific heat results show no sign of long-range magnetic order down to $\sim$0.1~K either, but only a Schottky anomaly that is continuously mediated by the external magnetic fields. Additionally, inelastic neutron scattering is employed to detect low-energy spin excitations in \ryvo. The absence of magnetic excitation signals as well as static magnetic order down to 97~mK aligns with the results from magnetic susceptibility and specific heat. Collectively, these findings point to a quantum disordered ground state with persistent spin dynamics, reminiscent of QSL behaviors. Our work provides a promising platform for further exploration of quantum magnetism in this new disorder-free system.
\end{abstract}

\maketitle

\section{Introduction}

In triangular-lattice rare-earth compounds, the interplay of multiple components such as geometrical frustration, strong spin-orbital coupling, and crystalline electric field~(CEF) effects, can result in a variety of novel quantum states, including spin ices~\cite{Castelnovo2012,Gingras_2014}, Berezinskii-Kosterlitz-Thouless phase~\cite{nc11_1111,nc11_5631}, quantum spin liquids~(QSLs)~\cite{nature464_199,RevModPhys.89.025003,PhysRevB.94.035107,Broholmeaay0668}, etc. Notably, QSLs represent a highly entangled quantum state but with no long-range magnetic order even down to zero temperature~\cite{Anderson1973153,nature464_199}. In a QSL, the elementary excitations are fractionalized quasiparticles, such as $S$=1/2 spinons~\cite{nature464_199,imai2016quantum,RevModPhys.89.025003,Broholmeaay0668}, in stark contrast with the collective excitation mode of integer-spin magnons in conventional magnets~\cite{PhysRev.102.1217,RevModPhys.90.015005}. These unique properties make QSLs promising candidates for fault-tolerant quantum computation~\cite{Kitaev20032,aop321_2,RevModPhys.80.1083,Barkeshli722}. Thus, the search for real QSL materials has become a topical issue in the field of strongly correlated electrons in recent years.

Among the proposed QSL candidates, the kagome-lattice compound ZnCu$_3$(OH)$_6$Cl$_2$ (also known as herbertsmithite) is a celebrated one~\cite{RevModPhys.88.041002,RevModPhys.89.025003,Broholmeaay0668}. While earlier inelastic neutron scattering~(INS) experiments reveal a broad continuum of magnetic spectra~\cite{nature492_406}, which was ever attributed to the gapless excitations of itinerant spinons expected in a QSL, subsequent investigations are at odds with this claim~\cite{science350_655,PhysRevB.94.060409,PhysRevLett.127.267202,np17_1109}. The central debate lies in whether antisite disorder plays a significant role in modulating its low-energy magnetic excitations~\cite{PhysRevLett.127.267202,PhysRevB.106.174406}, since there are approximately 15$\%$ of excess Cu$^{2+}$ impurities occupying out-of-plane nonmagnetic Zn$^{2+}$ sites and it would obscure the intrinsic physics~\cite{doi:10.1021/ja1070398,RevModPhys.88.041002}. Aside from that, the triangular-lattice rare-earth compound YbMgGaO$_4$ has also garnered tremendous attention as a QSL candidate. Although quite a few measurements show possible QSL behaviors~\cite{sr5_16419,prl115_167203,PhysRevLett.117.097201,nature540_559,np13_117,nc8_15814}, later findings challenge this scenario, including the absence of magnetic thermal conductivity~\cite{PhysRevLett.117.267202} and the occurrence of a frequency-dependent peak in ac magnetic susceptibility~\cite{PhysRevLett.120.087201,nc12_4949}. Considering the inherent site mixing between nonmagnetic Mg$^{2+}$ and Ga$^{3+}$ ions is striking in this compound~\cite{sr5_16419,prl115_167203}, it is suggested that the ground state is more likely characterized as a spin glass~\cite{PhysRevLett.120.087201}. These examples highlight how structural disorder hinders the direct investigation of the intrinsic ground state, emphasizing the demand for disorder-free QSL candidates. In this regard, the optimized NaYb$Ch_2$ ($Ch$ = O, S, Se) compounds have emerged as promising candidates~\cite{Liu_2018,doi:10.1021/acs.chemrev.0c00641,np15_1058}. Actually, various types of QSLs have been theoretically proposed to date, such as the QSL with spinon Fermi surface~\cite{PhysRevB.74.014408,nature540_559,PhysRevX.11.021044}, the Dirac QSL~\cite{PhysRevLett.102.047205,PhysRevLett.123.207203}, the $Z_{\rm 2}$ gapped QSL~\cite{PhysRevB.94.121111,PhysRevB.92.041105}, etc., it is crucial to realize these theoretical models with realistic material systems.

Recently, a new ytterbium-based triangular-lattice compound \kyvo was reported~\cite{doi:10.1021/ic301922e,PhysRevB.104.144411}. Due to the significant difference in ionic radii, structural disorder is prohibited in this material~\cite{PhysRevB.104.144411}. In \kyvo, magnetic Yb$^{3+}$ ions with effective spin $J_{\rm eff}$ = 1/2 form perfect triangular layers that are well separated along the crystallographic $c$ axis~\cite{doi:10.1021/ic301922e,PhysRevB.104.144411}. Magnetic susceptibility and specific heat results show neither long-range magnetic order nor spin freezing down to 0.5~K~\cite{PhysRevB.104.144411}. Electronic-structure calculations reveal strong spin-orbit coupling and considerable magnetocrystalline anisotropy in this regime, suggesting the potential for QSL physics~\cite{PhysRevB.104.144411}. Simultaneously, systematic research on the QSL candidates NaYb$Ch_2$ has demonstrated that substituting Na with other alkali-metal elements effectively tunes the strength of exchange interactions along different paths, resulting in distinct quantum states in the phase diagram~\cite{npjqm8_48,np20_74,scheie2024spectrumlowenergygaptriangular}. As a consequence, exploring such a substitution effect in the new compound \kyvo is highly desirable.

In this work, we substituted K with Rb and Cs in \kyvo, and synthesized two new triangular-lattice compounds \ryvo and \cyvo. X-ray diffraction~(XRD) characterizations reveal higher phase purity and larger separation between magnetic layers in our samples compared to \kyvo. Magnetic susceptibility measurements show that neither of the compounds undergoes a phase transition down to 1.8~K. A fit to the low-temperature data yields the Curie-Weiss temperature of -1.40~K and -0.43~K for \ryvo and \cyvo, respectively, indicating the dominant antiferromagnetic interactions. Moreover, the absence of bifurcation between zero-field-cooling~(ZFC) and field-cooling~(FC) data rules out the possibility of spin freezing. Specific heat measurements down to as low as $\sim$0.1~K show no $\lambda$-shaped peak indicative of a phase transition, but rather a Schottky anomaly that is continuously modulated by the applied magnetic fields. Due to the hygroscopic nature of \cyvo, INS experiments were performed on \ryvo. No signal from low-energy magnetic excitations is observed, which is consistent with the results of specific heat, where no density of states resulting from spinons expected in a QSL or magnons in a magnetically ordered state is captured. These findings suggest persistent spin fluctuations in the ground state of both compounds.

\section{Experimental Details}

Polycrystalline samples of \ryvo and \cyvo were synthesized by the conventional solid-state reaction method. The raw materials Rb$_2$CO$_3$~(99.99\%), Cs$_2$CO$_3$~(99.99\%), Yb$_2$O$_3$~(99.99\%), and V$_2$O$_5$~(99.99\%) were weighed with a stoichiometric proportion and thoroughly ground in an agate mortar. Then the well-mixed precursors were loaded into the alumina crucibles and heated at 750~$^\circ$C in air for 72 hours. To obtain pure-phase compounds, the intermediate grindings for twice were required.

XRD data were collected at room temperature in an X-ray diffractometer (SmartLab SE, Rigaku) using the Cu-$K_\alpha$ edge with a wavelength of 1.54~\AA. In the measurements, the scan angle of 2$\theta$ ranged from 15$^\circ$ to 80$^\circ$ with a step of 0.02$^\circ$ and a rate of 10$^\circ$/min. Rietveld refinements on the XRD data were performed by GSAS software. The dc magnetic susceptibility was measured on powder samples within 1.8-350~K using a Quantum Design physical property measurement system~(PPMS, Dynacool), equipped with a vibrating sample magnetometer option. Specific heat was measured by the relaxation method in a PPMS Dynacool equipped with a dilution refrigerator, allowing the measurements down to $\sim$0.1~K. INS experiments were conducted on FOCUS, a time-flight spectrometer at Paul Scherrer Institute~(PSI) at Villigen, Switzerland. About 5-g powder samples of \ryvo were loaded into a pure copper can and measured in a dilution refrigerator. It was able to cool down to a base temperature of 97~mK. The incident neutron energy was set to $E_{\rm i}$ = 3.69~meV with an energy resolution of $\Delta E$ = 0.05~meV (half width at half maximum, HWHM). For each temperature, we collected data for 15 hours.

\section{Results}
\subsection{Crystal structure and X-ray diffraction}

\begin{figure}[htb]
  \centering
  \includegraphics[width=0.98\linewidth]{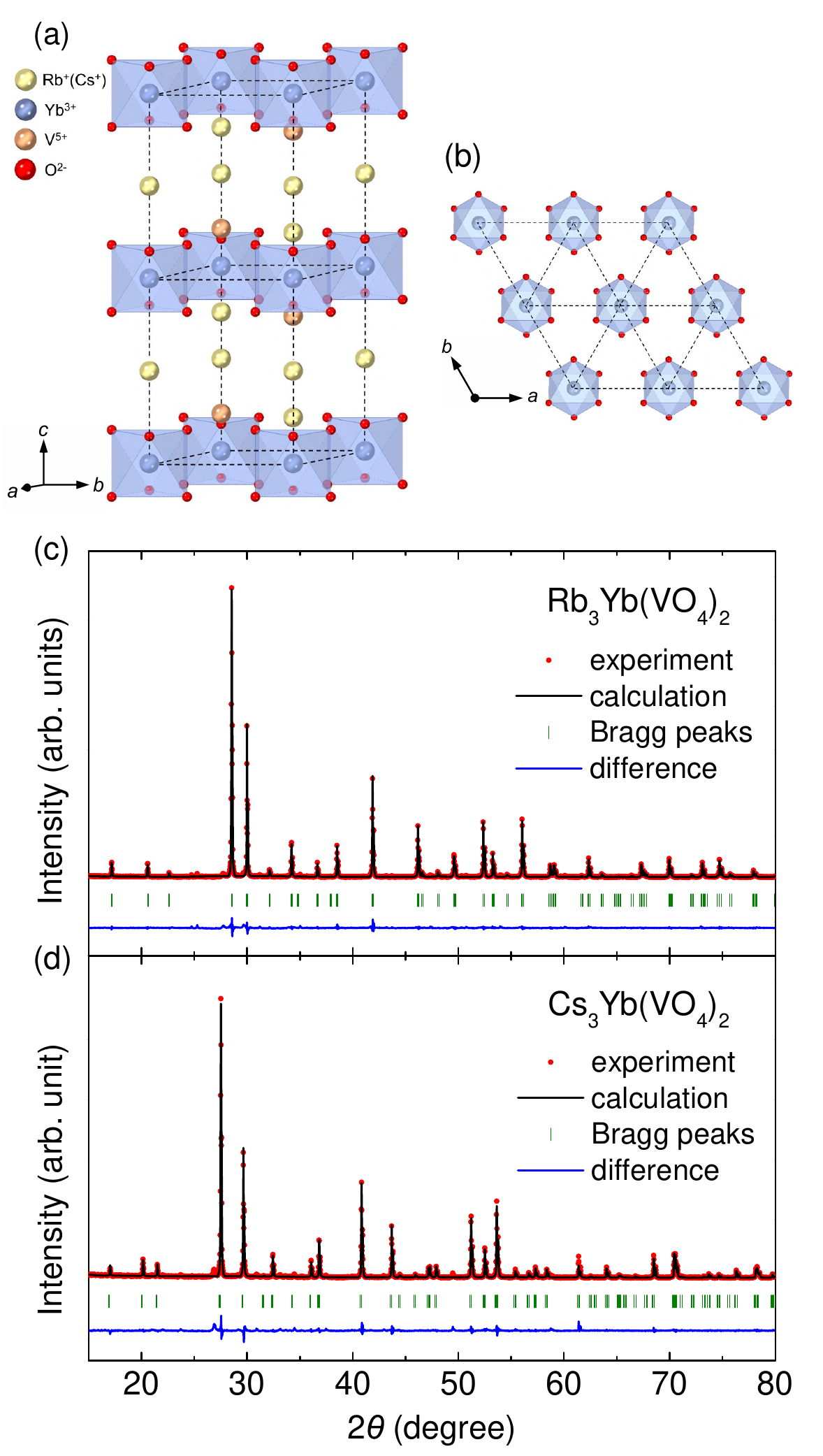}
  \caption{
  (a) Schematic crystal structure of \ryvo and \cyvo. (b) Top view of the magnetic triangular layer consisting of YbO$_6$ octahedra. (c) and (d) Rietveld refinement results of room-temperature XRD data for \ryvo and \cyvo, respectively. Red circles, black solid line, and green ticks indicate the measured data points, calculated results upon Rietveld refinements, and Bragg peak positions, respectively. Blue solid line represents the difference between the experiments and calculations.}
  \label{fig1}
\end{figure}

Figures~\ref{fig1}(a) and \ref{fig1}(b) depict the schematics of the crystal structure and a magnetic triangular layer of \ryvo and \cyvo, respectively. Both compounds share the same space group $P\bar3m$1 as \kyvo~\cite{doi:10.1021/ic301922e,PhysRevB.104.144411}. The magnetic triangular layers consist of edge-sharing YbO$_6$ octahedra and they have a simple $AAA$ stacking arrangement along the crystallographic $c$ axis, which is similar to that of the cobalt-based QSL candidate Na$_2$BaCo(PO$_4)_2$~\cite{Zhong14505}. Nonmagnetic ions are alternately interposed among these magnetic layers that minimizes the interlayer magnetic exchange interactions, featuring a quasi-two-dimensional magnetism.

Figures~\ref{fig1}(c) and \ref{fig1}(d) show the room-temperature XRD patterns along with the Rietveld refinement results of these two compounds. The refinements were carried out using the structural model of \kyvo~\cite{PhysRevB.104.144411}. As can be seen, there is a good agreement between the experimental and calculated results, with no impurity phase detected in either compound. For \ryvo, it gives rise to the lattice parameters $a = b$ = 5.95632(25)~\AA, $c$ = 7.86102(33)~\AA, and $\alpha$ = $\beta =$ 90$^\circ$, $\gamma$ = 120$^\circ$. For \cyvo, they are $a = b$ = 6.04659(6)~\AA, $c$ = 8.30327(8)~\AA, and $\alpha$ = $\beta =$ 90$^\circ$, $\gamma$ = 120$^\circ$. The detailed structural information is summarized in Table~\ref{tab:para}. The continuous increase in lattice constants from \kyvo to \cyvo proves that substituting K with Rb and Cs enlarges the unit cell, as expected due to the larger ionic radii of the latter two alkali-metal elements. Next, we consider the ratio of interlayer to intralayer distance $d_{\rm inter}/d_{\rm intra}$ for the magnetic Yb$^{3+}$ ions. It increases from 1.29 for \kyvo to 1.37 for \cyvo. This means such a substitution with larger alkali-metal ions can enhance the separation of magnetic ions along the $c$ axis and it further weakens the interlayer magnetic interactions. Notably, the value of $d_{\rm inter}/d_{\rm intra}$ herein is comparable to that of 1.32 in Na$_2$BaCo(PO$_4)_2$~\cite{Zhong14505}, a well-known triangular QSL candidate, supporting the two-dimensional magnetism in this new material system. The final refinement parameter of goodness of fitting $\chi^2$ is 4.353 and 3.654 for these two compounds, respectively, which is comparable to that of 4.1 for \kyvo~\cite{PhysRevB.104.144411}, confirming the high purity of our synthesized samples and the high quality of our refinements. Additionally, the atomic occupancies in these compounds are all 1, evidencing the disorder-free nature in structure.

\begin{table*}[htb]
  \begin{threeparttable}
\caption{Main structural parameters extracted from XRD pattern refinements for $A_{\rm 3}$Yb(VO$_4)_3$ ($A$ = K, Rb, Cs).}
\label{tab:para}
\begin{tabular*}{\textwidth}{@{\extracolsep{\fill}}ccccccccccc}
\hline
\hline
\begin{minipage}{2cm}\vspace{1mm} Compound \vspace{1mm} \end{minipage} & Atom & Wyckoff position & $x$ & $y$ & $z$ & Occupancy & $a$~(\AA) & $c$~(\AA) & $d_{\rm inter}/d_{\rm intra}$ & $\chi^2$\\
\hline
\begin{minipage}{2cm}\vspace{1mm}  \vspace{1mm} \end{minipage} & K(1) & 2$d$ & 0.6667 & 0.3333 & 1.2961 & 1 & & & & \\
  & K(2) & 1$a$ & 0 & 0 & 1 & 1 & & & & \\
 \kyvo & Yb & 1$b$ & 1 & 0 & 0.5 & 1 & 5.85 & 7.58 & 1.29 & 4.1~\cite{PhysRevB.104.144411}\\
  & V & 2$d$ & 0.6667 & 0.3333 & 0.7474 & 1 & & & & \\
  & O(1) & 2$d$ & 0.6667 & 0.3333 & 0.9595 & 1 &  &  &  &  \\
 & O(2) & 6$i$ & 0.3499 & 0.1750 & 0.6693 & 1 & & & & \\
\hline
\begin{minipage}{2cm}\vspace{1mm}  \vspace{1mm} \end{minipage} & Rb(1) & 2$d$ & 0.6667 & 0.3333 & 1.2877 & 1 & & & & \\
  & Rb(2) & 1$a$ & 0 & 0 & 1 & 1 & & & & \\
 \ryvo & Yb & 1$b$ & 1 & 0 & 0.5 & 1 & 5.96 & 7.86 & 1.32 & 4.4\\
  & V & 2$d$ & 0.6667 & 0.3333 & 0.7374 & 1 & & & & \\
  & O(1) & 2$d$ & 0.6667 & 0.3333 & 0.9470 & 1 &  &  &  &  \\
 & O(2) & 6$i$ & 0.3867 & 0.1934 & 0.6658 & 1 & & & & \\
\hline
\begin{minipage}{2cm}\vspace{1mm}  \vspace{1mm} \end{minipage} & Cs(1) & 2$d$ & 0.6667 & 0.3333 & 1.2655 & 1 & & & & \\
  & Cs(2) & 1$a$ & 0 & 0 & 1 & 1 & & & & \\
 \cyvo & Yb & 1$b$ & 1 & 0 & 0.5 & 1 & 6.05 & 8.30 & 1.37 & 3.7\\
  & V & 2$d$ & 0.6667 & 0.3333 & 0.7193 & 1 & & & & \\
  & O(1) & 2$d$ & 0.6667 & 0.3333 & 0.9310 & 1 &  &  &  &  \\
 & O(2) & 6$i$ & 0.3970 & 0.1985 & 0.6650 & 1 & & & & \\
\hline
\hline
\end{tabular*}
\end{threeparttable}
\end{table*}

\subsection{Magnetic susceptibility}

\begin{figure}[htb]
  \centering
  \includegraphics[width=0.98\linewidth]{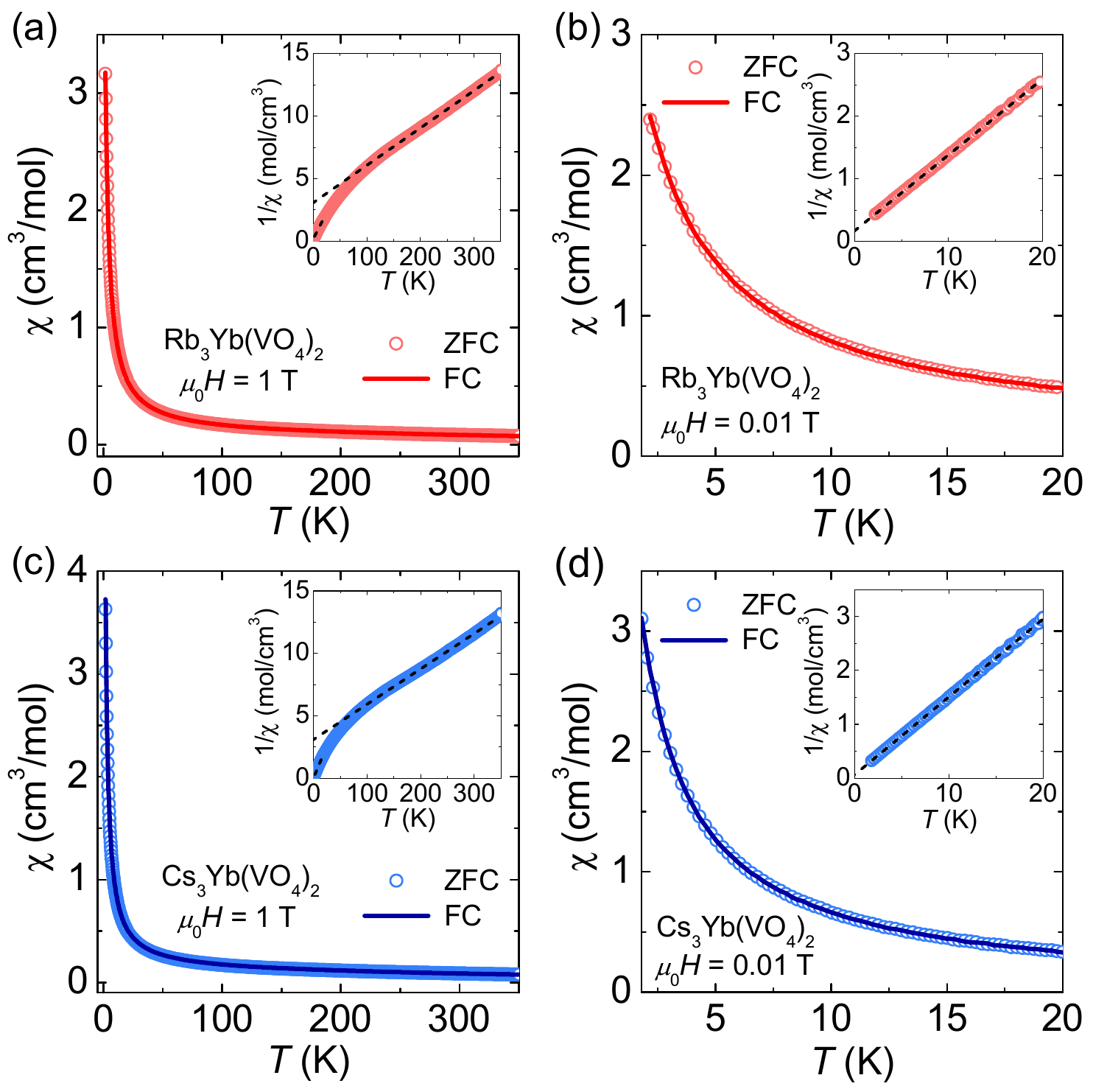}
  \caption{
  (a) and (c) Temperature dependence of magnetic susceptibility~($\chi$) for \ryvo and \cyvo measured at 1~T, repectively. (b) and (d) The low-temperature magnetic susceptibility results measured at a small field of 0.01~T. Insets show the corresponding inverse susceptibility data for these two compounds. The dashed lines are the fits with the Curie-Weiss law. All the data were collected in both zero-field-cooling~(ZFC) and field-cooling~(FC) conditions.}
  \label{fig2}
\end{figure}

We then characterized the magnetic properties of both compounds by measuring their magnetic susceptibility~($\chi$), which was obtained by using the magnetization~($M$) values divided by the applied magnetic fields~($\mu_{\rm 0}H$). Figures~\ref{fig2}(a) and \ref{fig2}(c) present the temperature dependence of magnetic susceptibility results for \ryvo and \cyvo measured at 1~T, respectively. There is no indication of a magnetic phase transition over a broad temperature range of 1.8-350~K, suggesting the persistent paramagnetic behavior down to low temperatures. The insets show the inverse susceptibility data for these two compounds. At high temperatures, the data display a linear behavior and it can be fitted by the Curie-Weiss law $\chi$ = $C/(T-\Theta_{\rm CW}$), where $C$ and $\Theta_{\rm CW}$ denote the Curie-Weiss constant and the Curie-Weiss temperature, respectively. From a fit within 150 to 340~K, we obtain $\Theta_{\rm CW}$ = -105.68 and -114.83~K, and the effective moment $\mu_{\rm eff}$ = 4.63 and 4.78~$\mu_{\rm B}$ for \ryvo and \cyvo, respectively. The yielded values of the effective moment are in good agreement with that of 4.54~$\mu_{\rm B}$ for a free Yb$^{3+}$ ion in the paramagnetic phase~\cite{PhysRevB.99.180401}. However, the high-temperature fit does not capture the physics associated with the magnetic ground state, since electrons would be thermally populated to the excited CEF levels at high temperatures due to the CEF effect. This is confirmed by the deviation from the linear behavior as temperature decreases below 100~K. To better understand the intrinsic physics of the magnetic ground state and minimize the influence of the applied magnetic field, we conducted low-temperature susceptibility measurements at a small field of 0.01~T. The results are exhibited in Figs.~\ref{fig2}(b) and \ref{fig2}(d). Neither anomaly nor bifurcation between ZFC and FC conditions is observed, indicating the absence of a phase transition or spin freezing at low temperatures. The inverse susceptibility data follow well a temperature-linear behavior below 20~K, as shown in the insets of Figs.~\ref{fig2}(b) and \ref{fig2}(d). To exclude the extrinsic magnetic contributions, the Curie-Weiss fit was performed with the expression $\chi$ = $\chi_{\rm 0}$ + $C/(T-\Theta_{\rm CW}$), where $\chi_{\rm 0}$ denotes temperature-independent term. It gives rise to $\Theta_{\rm CW}$ = -1.40 and -0.43~K, $\mu_{\rm eff}$ = 2.29 and 2.10~$\mu_{\rm B}$, and $\chi_{\rm 0}$ = 9.86 and 9.02~$\times$~10$^{-2}$~cm$^3$/mol for \ryvo and \cyvo, respectively. These parameters summarized in Table~\ref{tab:suscep} are comparable to those of \kyvo~\cite{PhysRevB.104.144411}. The negative sign of $\Theta_{\rm CW}$ reflects that the magnetic ground state is dominated by antiferromagnetic interactions between Yb$^{3+}$ moments. We also extracted the $\Theta_{\rm CW}$ values from the low-temperature susceptibility data measured at 1~T, which are -1.54~K and -0.59~K for \ryvo and \cyvo, respectively. They are in good agreement with those obtained at the small field of 0.01~T, both suggesting the existence of antiferromagnetic interactions. On the mean-field level, the Curie-Weiss temperature can be expressed as $\Theta_{\rm CW}$ = -$zJS(S+1)$/3$k_{\rm B}$~\cite{PhysRevLett.116.157201,PhysRevB.107.064421}, where $z$, $k_{\rm B}$, and $J$ are the number of nearest-neighbor spins, the Boltzman constant, and the nearest-neighbor exchange interactions, respectively. From this relation, $J$ is estimated to be 0.93 and 0.28~K for \ryvo and \cyvo, respectively. Additionally, the reduced effective moments at low temperatures compared to that obtained by the high-temperature fit imply the development of spin correlations, which is also observed in other ytterbium-based QSL candidates, such as YbMgGaO$_4$ and NaYb$Ch_2$ family~\cite{sr5_16419,Liu_2018}.

\begin{table}[htb]
\caption{Curie-Weiss temperature~($\Theta_{\rm CW}$) and effective moment~($\mu_{\rm eff}$) yielded by the Curie-Weiss fitting to low-temperature~(LT) and high-temperature~(HT) data.}
\label{tab:suscep}
\begin{tabular*}{\columnwidth}{@{\extracolsep{\fill}}ccccc}
\hline\hline
\begin{minipage}{2cm}\vspace{1mm}\centering  Compound \vspace{1mm} \end{minipage} & $\Theta_{\rm CW}^{\rm LT}$ (K) & $\mu_{\rm eff}^{\rm LT}$ ($\mu_{\rm B}$)& $\Theta_{\rm CW}^{\rm HT}$ (K) & $\mu_{\rm eff}^{\rm HT}$ ($\mu_{\rm B}$)\\
\hline
\begin{minipage}{2cm}\vspace{1mm}\centering  \kyvo \vspace{1mm} \end{minipage}  & -1 & 2.41  &  -110  & 5.06~\cite{PhysRevB.104.144411} \\
\ryvo   & -1.40 & 2.29 & -105.68 & 4.63 \\
\cyvo   & -0.43 & 2.10 & -114.83 & 4.78 \\
\hline\hline
\end{tabular*}
\end{table}

\begin{figure}[htb]
  \centering
  \includegraphics[width=0.98\linewidth]{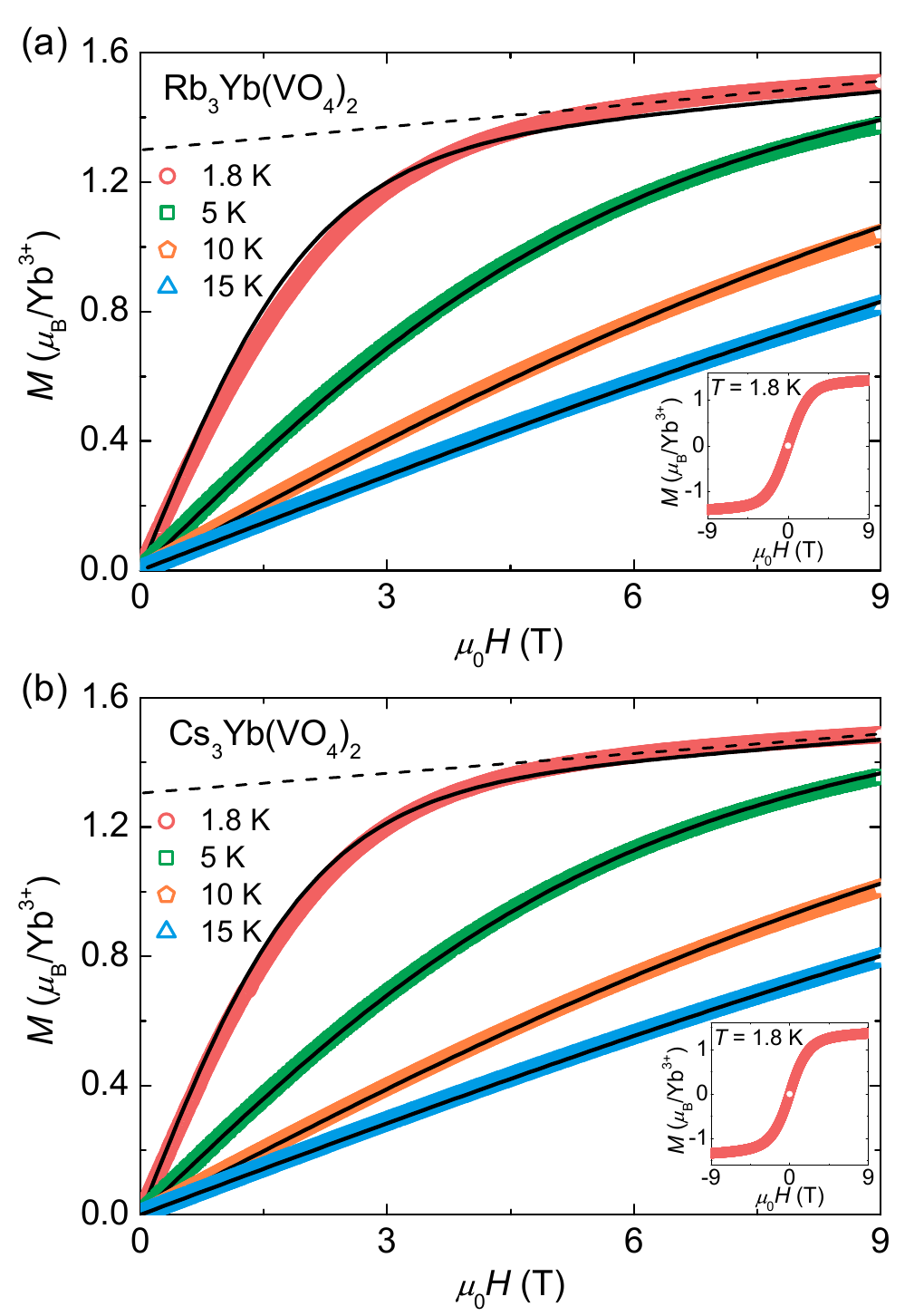}
  \caption{
  (a) and (b) Magnetic-field dependence of the magnetization measured at various temperatures for \ryvo and \cyvo, respectively. The solid lines are the fits to the data with the combination of Van Vleck contribution and the Brillouin function. The dashed lines denote the linear fits to the data for $\mu_{\rm 0}H\geq$ 7~T. The insets show magnetization loops from -9~T to 9~T measured at 1.8~K for \ryvo and \cyvo, respectively.}
  \label{fig3}
\end{figure}

To further reveal the magnetic properties of these two compounds, the measurements of isothermal field-dependent magnetization were performed at various temperatures, and the results are depicted in Fig.~\ref{fig3}. For \ryvo at 1.8~K, the magnetization increases linearly with the magnetic field to $\sim$1.5~T, followed by a smooth transition into a fully saturated regime above $\sim$6~T, where the magnetization curve reverts to a field-linear dependence again. The initial linear increase at low fields suggests the uniform magnetization of the compound, while it is stemmed from the extrinsic contribution of Van Vleck paramagnetic susceptibility~($\chi_{\rm VV}$) when the field is strong enough to fully polarize all the magnetic moments. Fitting the high-field data yields $\chi_{\rm VV}$ = 2.36~$\times$~10$^{-2}$~cm$^3$/mol and the saturated moment $M_{\rm sat}$ = 1.30~$\mu_{\rm B}$/Yb$^{3+}$. The Land\'{e} $g$ factor is calculated as 2.60 via the relationship $M_{\rm sat} = gJ_{\rm eff}\mu_{\rm B}$ when considering the effective spin $J_{\rm eff}$ = 1/2 that will be justified in the following specific heat part. For \cyvo, the magnetization behavior shows a similar trend as \ryvo, but with a slightly low saturation field of $\sim$5~T. This lower saturation field for \cyvo reflects the weaker spin interactions compared to \ryvo, consistent with the smaller value of $|\Theta_{\rm CW}|$ obtained in above magnetic susceptibility analysis. The fitting gives rise to $\chi_{\rm VV}$ = 2.03~$\times$~10$^{-2}$~cm$^3$/mol and $M_{\rm sat}$ = 1.31~$\mu_{\rm B}$/Yb$^{3+}$ for \cyvo, corresponding to $g$ = 2.62. As temperature increases, the magnetization continuously decreases and gradually deviates from the saturation behavior. It is due to the suppression of magnetic couplings between magnetic ions and the development of thermal activation of the excited CEF levels. The magnetization behavior at various temperatures can be quantitatively described by the function $M(H)$ = $\chi_{\rm VV}H$ + $gJ_{\rm eff}N_{\rm A}\mu_{\rm B}B_J(H)$~\cite{PhysRevB.107.064421}, where $N_{\rm A}$ and $B_J$ are respectively the Avogadro constant and the Brillouin function for an effective spin-1/2 system. It results in an average Land\'e factor $g$ = 2.57 for \ryvo and 2.58 for \cyvo, which is in perfect agreement with the values of 2.60 for \ryvo and 2.62 for \cyvo derived from saturated moments. Additionally, there is no hysteresis observed in the whole magnetization loops from -9~T to 9~T measured at 1.8~K for both compounds, as shown in the insets of Fig.~\ref{fig3}, consistent with the negative sign of the derived $\Theta_{\rm CW}$ values. They both indicate the dominant antiferromagnetic interactions in the ground state.

\subsection{Specific heat}

\begin{figure*}[htb]
  \centering
  \includegraphics[width=6.8in]{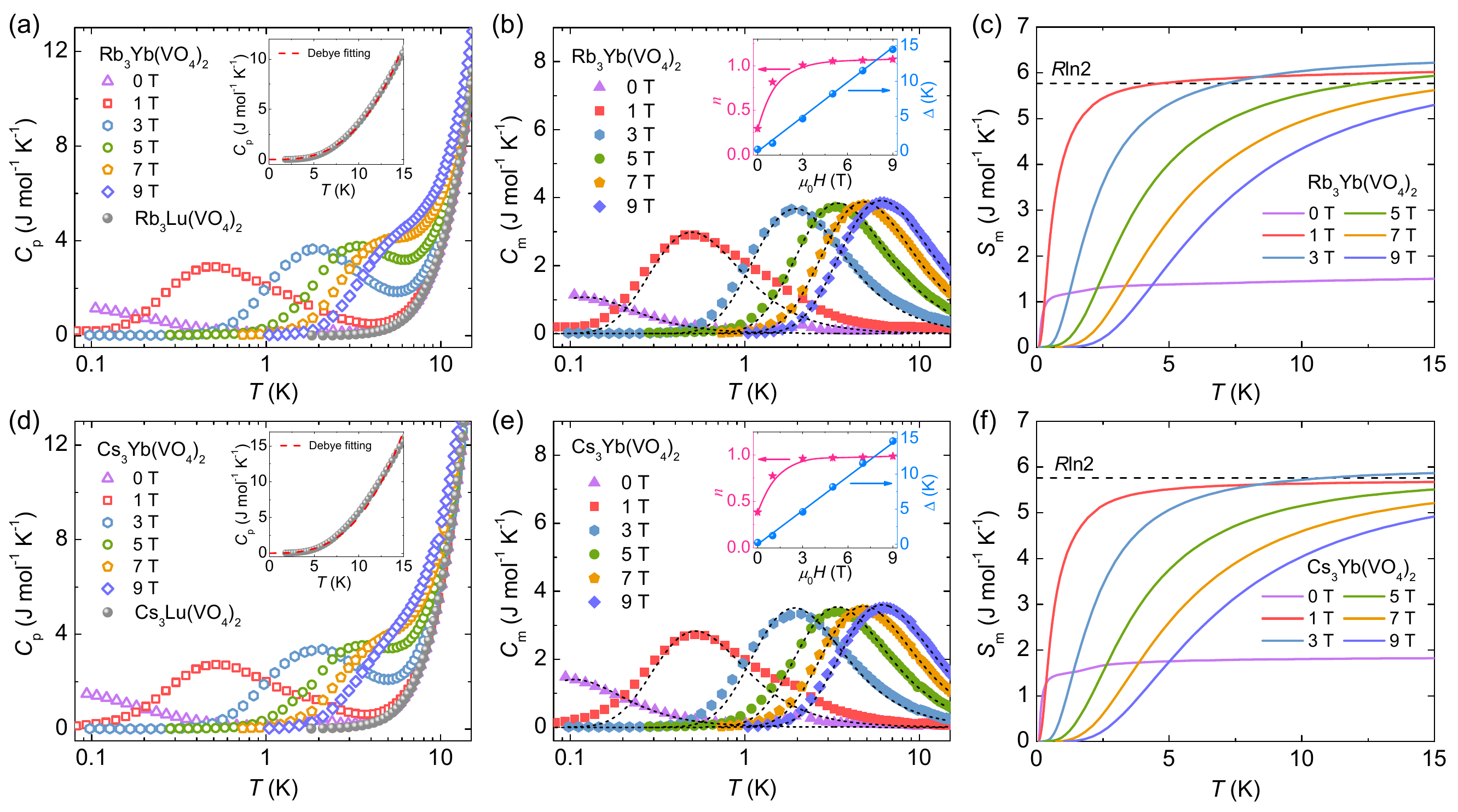}
  \caption{
  (a) and (d) Ultralow-temperature specific heat~($C_{\rm p}$) results of \ryvo and \cyvo at various magnetic fields, respectively. The insets respectively show the specific heat of nonmagnetic reference compounds Rb$_3$Lu(VO$_4)_2$ and Cs$_3$Lu(VO$_4)_2$, which can be nicely fitted by the Debye model as $C_{\rm p}\propto T^3$. (b) and (e) Magnetic specific heat ($C_{\rm m}$) of both compounds after subtracting the lattice contributions with the nonmagnetic isostructural compounds of Rb$_3$Lu(VO$_4)_2$ and Cs$_3$Lu(VO$_4)_2$. The dashed lines represent the fits with a two-level Schottky function, as described in the main text. The insets show the magnetic-field dependence of the percent of free spins $n$~(left vertical axis) and the energy gap $\Delta$~(right vertical axis). (c) and (f) Temperature dependence of magnetic entropy~($S_{\rm m}$) at various fields, which were obtained by integrating $C_{\rm m}/T$ data. The dashed lines indicate the value of $R$ln2, where $R$ is the ideal gas constant.}
  \label{fig4}
\end{figure*}

To further elucidate the nature of their magnetic ground states, we measured the specific heat~($C_{\rm p}$) down to a quite low temperature of $\sim$0.1~K. Figure~\ref{fig4}(a) shows the $C_{\rm p}$ results of \ryvo at various magnetic fields as well as a nonmagnetic isostructural compound Rb$_3$Lu(VO$_4)_2$ for comparison. At zero field, there is no sharp $\lambda$-type peak observed even at subkelvin temperatures, but an upturn below $\sim$1~K, ruling out the onset of long-range magnetic order. When an external magnetic field is applied on this compound, a broad peak emerges and shifts towards higher temperatures with the increasing fields. This behavior is reminiscent of the Schottky anomaly due to the Zeeman splitting of a Kramers doublet ground state. The similar phenomena is also observed in other geometrically frustrated ytterbium-based quantum magnets, such as KBaYb(BO$_3)_2$~\cite{PhysRevB.103.104412}, YbBO$_3$~\cite{PhysRevB.107.064421}, Ba$_6$Yb$_2$Ti$_4$O$_{17}$~\cite{PhysRevB.109.024427}, etc. As temperature increases, the $C_{\rm p}$ data of \ryvo and Rb$_3$Lu(VO$_4)_2$ are nearly overlapped above $\sim$10~K, suggesting a significant contribution from phonons. The inset of Fig.~\ref{fig4}(a) shows the specific heat of Rb$_3$Lu(VO$_4)_2$ individually, which is nicely fitted by a Dybye model as $C_{\rm p}\propto T^3$. This can be actually taken as the pure lattice part for \ryvo. For \cyvo, the specific heat depicted in Fig.~\ref{fig4}(d) follows a pattern akin to that of \ryvo, both pointing to a magnetically disordered ground state. The only difference is that the nonmagnetic analog for \cyvo is Cs$_3$Lu(VO$_4)_2$, which also fits well to the Debye model, as shown in the inset of Fig.~\ref{fig4}(d).

Figures~\ref{fig4}(b) and \ref{fig4}(e) display the magnetic specific heat~($C_{\rm m}$) results of these two compounds, obtained by subtracting the phonon contributions. No phase transition but a broad hump is observed, confirming the absence of long-range magnetic order. The $C_{\rm m}$ data at different fields are well fitted by a two-level Schottky function~\cite{PhysRevB.103.104412,PhysRevB.107.064421},
\begin{equation}\label{CEF}
  C_{\rm Sch}=nR(\frac{\Delta}{T})^2\frac{e^{-\frac{\Delta}{T}}}{(1+e^{-\frac{\Delta}{T}})^2},
\end{equation}
where $n$ is the molar fraction of free spins, $R$ is the gas constant, and $\Delta$ is the Zeeman gap between two levels in the Kramers doublet ground state. Taking $\Delta$ and $n$ as the fitting parameters, the resulting values for \ryvo are plotted as a function of $\mu_{\rm 0}H$ in the inset of Fig.~\ref{fig4}(b). It shows that $n$ increases rapidly with the increasing field and then saturates at the value of 1 when the field exceeds $\sim$3~T. This indicates that a fraction of spins are correlated at lower magnetic fields while all the free spins are excited at higher fields that contribute to the Schottky specific heat. For the gap $\Delta$, it increases linearly with the field as expected, since the Zeeman effect follows the function $\Delta = g\mu_{\rm B}\mu_{\rm 0}H$. A linear fit to the $\Delta$ values yields $g$ = 2.45, which is consistent with the values of 2.57 and 2.60 obtained by the above magnetic susceptibility analysis. For \cyvo, there is no significant difference in $n$ and $\Delta$ compared to \ryvo, as shown in the inset of Fig.~\ref{fig4}(e). The fit gives $g$ = 2.39 that also aligns well with 2.58 and 2.62 from the susceptibility results.

From the above magnetic specific heat results, we obtained magnetic entropy~($S_{\rm m}$) of both compounds by integrating the $C_{\rm m}/T$ data. Figure~\ref{fig4}(c) presents the temperature dependence of magnetic entropy at different fields for \ryvo. At zero field, $S_{\rm m}$ rapidly reaches a saturation status with the temperature increasing to $\sim$1~K. The saturated $S_{\rm m}$ value is 1.59~J~mol$^{-1}$~K$^{-1}$, which is only 28$\%$ of expected $R$ln2 for the $J_{\rm eff}$ = 1/2 magnetic ground state. This implies that the remaining $\sim$70$\%$ magnetic entropy would be released below 0.1~K due to the presence of strong spin correlations~\cite{PhysRevB.107.064421,PhysRevB.109.024427}. When an external field is applied on this material, $S_{\rm m}$ immediately recovers to $R$ln2 at 1~T, indicative of a two-level magnetic system. As the field continues to increase, the temperature at which magnetic entropy begins to saturate shifts to higher values, suggesting the broadening of the spin gap~\cite{PhysRevB.109.024427,PhysRevB.104.L220403,shu2020}. For \cyvo, the overall behavior of magnetic entropy is quite similar to that of \ryvo, as shown in Fig.~\ref{fig4}(f). The release of zero-field magnetic entropy above $\sim$0.1~K rises to 33$\%$ of $R$ln2. These results indicate that both compounds share the same quantum disordered ground state.

\subsection{INS spectra}

\begin{figure}[htb]
  \centering
  \includegraphics[width=0.98\linewidth]{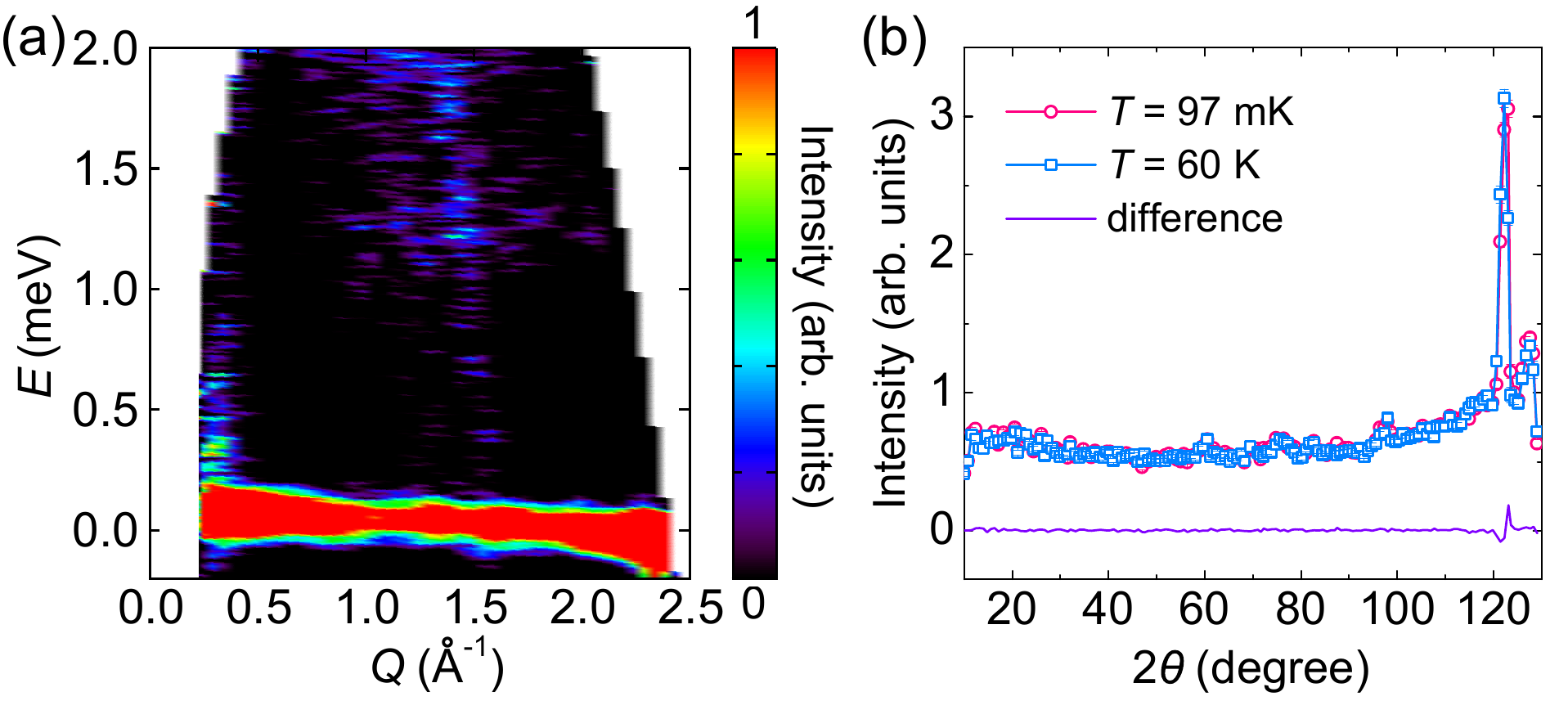}
  \caption{
  (a) Inelastic neutron scattering spectrum of \ryvo polycrystalline samples measured at 97~mK. (b) Elastic neutron scattering data at the lowest measured temperature of 97~mK and the highest temperature of 60~K, which were obtained by integrating the intensity in an energy window of [-0.05, 0.05]~meV.}
  \label{fig5}
\end{figure}

To further verify the characteristic of persistent spin dynamics in the ground state and detect any potential magnetic excitation signals, INS technique was employed with the spectrometer FOCUS that can cool the samples down to an ultralow temperature of 97~mK. Considering the hygroscopic nature of \cyvo during the sample preparation process, the polycrystalline samples of \ryvo were selected for the INS measurements. Figure~\ref{fig5}(a) shows the low-energy magnetic excitation spectrum collected at 97~mK. Surprisingly, apart from the high scattering intensity around zero energy, which is attributed to its crystal structure, no magnetic signals such as diffuse scattering or well-defined spin waves are captured below 2~meV, indicating the absence of gapless magnetic excitations for this compound. It should be noted that the weak signal with some sporadic intensity above 1~meV is not a manifestation of intrinsic magnetic properties, but rather the background of the spectrometer. Based on the magnetic spectra measured at various temperatures, the integrated intensities extracted from the elastic channels are plotted as a function of the diffraction angles in Fig.~\ref{fig5}(b). The diffraction data for the lowest measured temperature of 97~mK and the highest one of 60~K are nearly overlapped, with no additional magnetic Bragg peaks observed, especially in the low-angle range where magnetic peaks typically show up for a magnetically ordered system~\cite{prl98_107204,PhysRevB.102.134407}. This result affirms the absence of long-range magnetic order, consistent with the findings from above magnetic susceptibility and specific heat measurements.

\section{Discussions}

The discovery of YbMgGaO$_4$ has sparked the enormous interest in searching for QSLs in triangular-lattice ytterbium-based antiferromagnets~\cite{sr5_16419,prl115_167203,Liu_2018}. However, the consequent problem of structural disorder complicates the situation~\cite{RevModPhys.89.025003,Broholmeaay0668,ZhenMa:106101,npjqm4_12}. In this work, the newly synthesized compounds \ryvo and \cyvo eliminate the influence from disorder, thereby providing a clearer window into the intrinsic physics of the investigated system. Notably, there has been a relevant research on the isostructural compound \kyvo, which shows the similar magnetic behaviors~\cite{PhysRevB.104.144411}. Electronic-structure calculations reveal significant magnetic anisotropy, which is of easy-plane type and favorable to the realization of a QSL~\cite{PhysRevB.104.144411}. In particular, the magnetization results from the single crystal sample of an analogue Rb$_3$Yb(PO$_4$)$_2$ have provided the direct evidence for it, where the magnetic susceptibility in the $ab$ plane is significantly larger than that out of the plane~\cite{shu2020}. We believe this should be true for \ryvo and \cyvo as well when considering the isostructural characteristic and comparable lattice constants compared with \kyvo and Rb$_3$Yb(PO$_4$)$_2$. In contrast, the higher ratio of interlayer to intralayer distance between magnetic Yb$^{3+}$ ions in \cyvo~(1.37) and \ryvo~(1.32) compared to \kyvo~(1.29) renders them closer to an ideal two-dimensional magnetic system. This means stronger magnetic anisotropy can be expected in the newly synthesized materials~\cite{PhysRevB.94.035107,PhysRevX.9.021017}. Moreover, substituting K with heavier Rb and Cs elements in this material family further promises the crystal perfection of the prepared samples~\cite{Liu_2018,PhysRevX.11.021044,PhysRevB.108.174432}. It is a crucial point especially for the magnetic system with weak magnetic interactions less than 1~K~\cite{ZhenMa:106101,npjqm4_12}. As such, \ryvo and \cyvo offer a more suitable platform to study the nontrivial physics. While currently the smoking-gun evidence for a QSL remains elusive~\cite{RevModPhys.89.025003,Broholmeaay0668,imai2016quantum,ZhenMa:106101,npjqm4_12}, the absence of long-range magnetic order down to $\sim$0.1~K, one order of magnitude lower than the interaction energy scale of $|\Theta_{\rm CW}|\sim$~1.4~K especially for \ryvo, strongly supports a quantum disordered ground state. Considering both the sample stability and the strength of spin interactions, \ryvo stands out for the most promising candidate for future investigations among these three compounds.

Based on these results, we now propose several possibilities for their exact ground states. 1) The most natural idea is a gapped QSL. The absence of gapless magnetic excitations in INS spectra as well as ultralow-temperature specific heat seems to support this scenario. However, the weak spin interactions of $\sim$1~K may not be able to generate a too large gap to be detected by the low-energy INS measurements below 2~meV. Further investigations using higher-energy neutrons are necessary to conclusively determine the presence of the spin gap. 2) It may be a gapless QSL. Compared with the extensively studied QSL candidate YbMgGaO$_4$, where the distance between the nearest-neighbor Yb$^{3+}$ ions is $\sim$3.4~\AA, resulting in a Curie-Weiss temperature $\Theta_{\rm CW}\sim$~-4~K~\cite{sr5_16419,prl115_167203}, it is reasonable to expect that the relatively large Yb-Yb distance of $\sim$6~\AA~in these two compounds would lead to weaker magnetic interactions. Consequently, the magnetic excitation signals are confined to a quite small energy scale, as observed in another QSL candidate Ce$_2$Zr$_2$O$_7$ with weak spin interactions ($\Theta_{\rm CW}$ = -0.57~K), in which the spectral weight of magnetic excitations is predominantly distributed below 0.2 meV and convoluted into the elastic signals~\cite{np15_1052}. This thus calls for further inelastic neutron scattering investigations with higher energy resolution. Moreover, the determination of the CEF scheme via high-energy neutrons will help to assess the distribution of low-energy magnetic spectral weight and thus elucidate the nature of the ground state. 3) It is a quantum dipole state, as proposed for another triangular-lattice ytterbium-based compound Ba$_3$Yb(BO$_3)_3$~\cite{PhysRevB.104.L220403,PhysRevB.106.014409}. In this case, the dominant component of magnetic couplings is dipole-dipole interactions rather than superexchange interactions~\cite{np14_405,PhysRevX.12.021015}. In fact, the magnetic behaviors including the small value of $\Theta_{\rm CW}$, the absence of magnetic transition and gapless magnetic excitations determined by specific heat and INS, all exhibit the striking similarities to this scenario. Further confirmation especially upon single-crystal samples is required for this possibility. 4) It is a valence bond solid, where two nearest-neighbor spins form a singlet, prohibiting the establishment of long-range magnetic order even at low temperatures~\cite{nc9_4367,PhysRevX.8.031028,PhysRevX.8.041040,nc10_2561}. This scenario was previously proposed for YbMgGaO$_4$~\cite{nc9_4367,PhysRevX.8.031028}. The key difference is that strong Mg$^{2+}$-Ga$^{3+}$ disorder in YbMgGaO$_4$ disrupts the well-arranged singlets and induces the low-energy magnetic excitations~\cite{PhysRevX.8.031028}, while the disorder-free nature for \ryvo and \cyvo preserves these singlets and thus prevents the gapless collective excitations. Whatever which case it is, they all exhibit the fascinating physics that will attract significant attention to this fertile ground in the future. Moreover, the weak spin interactions in this system enable facile external tuning. Notably, the substantial magnetic entropy change with the small applied fields for both compounds suggest potential applications in adiabatic demagnetization refrigeration at sub-kelvin temperatures, like the behaviors observed in another geometrically frustrated Yb-based antiferromagnet Yb$_2$Be$_2$GeO$_7$~\cite{PhysRevB.110.144445}.

\section{Summary}

In summary, we have successfully synthesized two new triangular-lattice rare-earth compounds \ryvo and \cyvo. XRD characterizations reveal that the samples are of high crystallization quality, with no structural disorder detected. Magnetic susceptibility measurements down to 1.8~K show no sign of magnetic phase transition or spin freezing. Furthermore, ultralow-temperature specific heat and INS experiments down to $\sim$0.1~K confirm the absence of long-range magnetic order. Notably, no contributions from gapless magnetic excitations, but only a Schottky anomaly is observed in specific heat. These findings point to a quantum disordered ground state with strong spin fluctuations, highlighting the nontrivial physics in this pristine geometrically frustrated system.

\section{Acknowledgements}

This work was supported by the National Key Projects for Research and Development of China with Grants No.~2021YFA1400400 and No.~2024YFA1409200, National Natural Science Foundation of China with Grants No.~12204160, No.~12225407, No.~12434005, No.~12074174, No.~12074111, No.~12204159, No.~12304119, and No.~52272108, Natural Science Foundation
of Jiangsu Province with Grant No.~BK20233001, Hubei Provincial Natural Science Foundation of China with Grants No.~2025AFB738 and No.~2023AFA105, Young Top-notch Talent Cultivation Program of Hubei Province, and Fundamental Research Funds for the Central Universities. We thank the neutron beam time from PSI with Proposal No.~20212437 and the support from Sample Environment Group for setting up the dilution refrigerator on FOCUS.

%

\end{document}